\documentstyle[preprint,pre,aps]{revtex}

\tighten

\begin{document}
\preprint{Submitted to Physical Review E.}

\title{Algorithmic information and simplicity in statistical physics}

\author{R\"udiger Schack\cite{MileEnd}}

\address{Center for Advanced Studies,
 Department of Physics and Astronomy,\\
 University of New Mexico, Albuquerque, NM 87131--1156}

\date{September 3, 1994}

\maketitle

\begin{abstract}
Given a list of $N$ states with probabilities $0<p_1\leq\cdots\leq
p_N$, the average conditional algorithmic information $\bar I$ to
specify one of these states obeys the inequality $H\leq\bar I<H+O(1)$,
where $H=-\sum p_j\log_2p_j$ and $O(1)$ is a computer-dependent
constant. We show how any universal computer can be slightly modified
in such a way that the inequality becomes $H\leq\bar I<H+1$, thereby
eliminating the computer-dependent constant from statistical physics.
\end{abstract}

\narrowtext

\section{INTRODUCTION}

Algorithmic information
theory\cite{Solomonoff1964,Kolmogoroff1965,Chaitin1987a}, in
combination with Landauer's principle\cite{Landauer1961,Landauer1988},
which specifies the unavoidable energy cost $k_BT\ln2$ for the erasure
of a bit of information in the presence of a heat reservoir at
temperature $T$, has been applied successfully to a range of problems:
the Maxwell demon paradox\cite{Bennett1982}, a consistent Bayesian
approach to statistical
mechanics\cite{Zurek1989a,Zurek1989b,Caves1993a,Caves1993b}, a
treatment of irreversibility in classical Hamiltonian chaotic
systems\cite{Caves1993b,Schack1992a}, and a characterization of
quantum chaos relevant to statistical
physics\cite{Caves1993b,Schack1993e,Schack1994b}. The algorithmic
information for a physical state is defined as the length in bits of
the shortest self-delimiting program for a universal computer that
generates a description of that
state\cite{Zurek1989b,Caves1990c}. Algorithmic information with
respect to two different universal computers differs at most by a
computer-dependent constant\cite{Chaitin1987a}. Although typically the
latter can be neglected in the context of statistical physics, the
presence of an arbitrary constant in a physical theory is
unsatisfactory and has led to criticism\cite{Denker1993}. In the
present paper, we show how the computer-dependent constant can be
eliminated from statistical physics.

In the following paragraphs we give a simplified account of the role
of algorithmic information in classical statistical physics. A more
complete exposition including the quantum case can be found in
Refs.~\cite{Zurek1989b,Caves1993b}.  We adopt here the
information-theoretic approach to statistical physics pioneered by
Jaynes\cite{Jaynes1983}. In this approach, the {\it state\/} of a
system represents the observer's knowledge of the way the system was
prepared. States are described by probability densities in phase
space; observers with different knowledge assign different states to
the system. Entropy measures the information missing toward a complete
specification of the system.

Consider a set of $N$ states ($N\geq2$) labeled by $j=1,\ldots,N$, all
having the same energy and entropy. The restriction to states of the
same energy and entropy is not essential, but it simplifies the
notation. Initially the system is assumed to be in a state in which
state $j$ is occupied with probability $p_j>0$. We assume throughout
that the states $j$ are labeled such that $0<p_1\leq\cdots\leq p_N$.
If an observation reveals that the system is in state $j$, the
increased knowledge is reflected in an entropy decrease $\Delta
S=-k_B\ln2\,H$ where $H=-\sum p_j\log_2p_j>0$ is
the original missing information measured in bits. To
make the connection with thermodynamics, we assume that there is a
heat reservoir at temperature $T$ to which all energy in the form of
heat must eventually be transferred, possibly using intermediate steps
such as storage at some lower temperature.  In the presence of this
fiducial heat reservoir, the entropy decrease $\Delta S$ corresponds
to a free energy increase $\Delta F=-T\Delta S=+k_BT\ln2\,H$.
Each bit of missing
information decreases the free energy by the amount $k_BT\ln2$; if
information is acquired about the system, free energy increases.

The fact that entropy can decrease through observation---which
underlies most proposals for a Maxwell demon---does not conflict with
the second law of thermodynamics because the observer's physical state
changes as a consequence of his interaction with the
system. Szilard\cite{Szilard1929} discovered that no matter how
complicated is the change in the observer's physical state, the
associated irreducible thermodynamic cost can be described solely in
terms of information. He found that in the presence of a heat
reservoir at temperature $T$ each bit of information acquired by the
observer has an energy cost at least as big as $k_BT\ln2$.
Total available work is reduced not only by
missing information, but also by
information the observer has acquired about the system. The physical
nature of the cost of information was clarified by
Bennett\cite{Bennett1982}, who applied Landauer's
principle\cite{Landauer1961,Landauer1988} to the Maxwell demon problem
and showed that the energy cost has to be paid when information is
erased.

To keep the Landauer erasure cost of the observational record as low
as possible, the information should be stored in maximally compressed
form. The concept of a maximally compressed record is formalized in
algorithmic information theory\cite{Chaitin1987a}.
Bennett\cite{Bennett1982} and Zurek\cite{Zurek1989a,Zurek1989b} gave
Szilard's theory its present form by using algorithmic information to
quantify the amount of information in an observational record.  In
particular, by exploiting Bennett's idea of a reversible
computer\cite{Bennett1982}, Zurek\cite{Zurek1989a} showed how an
observational record can be replaced by a compressed form at no
thermodynamic cost. This means that the energy cost of the
observational record can be reduced to the Landauer erasure cost of
the compressed form.

Let us denote by $s_j$ a binary string describing the $j$th state
($j=1,\ldots,N$). A detailed discussion of how a description of a
physical state can be encoded in a binary string is given
in\cite{Zurek1989b}. The exact form of the strings $s_j$ is of no
importance for the theory outlined here, however,
because the information needed
to generate a list of {\it all\/} the strings $s_j$ can be treated as
{\it background information}\cite{Caves1993b,Caves1990c}. Background
information is the information needed to generate a list
$s=((s_1,p_1),\ldots,(s_N,p_N))$ of all $N$ states together with their
probabilities; i.e., background information is the information the
observer has before the observation.

Algorithmic information is defined with respect to a specific
universal computer $U$. We denote by $I_U(s_j|s)$ the conditional
algorithmic information, with respect to the universal computer $U$,
to specify the $j$th state, given the background
information\cite{Chaitin1987a,Zurek1989b,Caves1990c}.  More precisely,
$I_U(s_j|s)$ is the length in bits of the shortest self-delimiting
program for $U$ that generates the string $s_j$, given a minimal
self-delimiting program to generate $s$.  For a formal definition of a
universal computer $U$ and of $I_U(s_j|s)$ see Sec.~\ref{opt}. It
should be emphasized that a minimal program that generates the list
$s$ of descriptions of all states and their probabilities can be short
even when a minimal program that generates the description $s_j$ of a
typical single state is very long\cite{Zurek1989b}.

Since total available work is reduced by $k_BT\ln2$ by each bit of
information the observer acquires about the system as well as by each
bit of missing information, the change in {\it total free energy\/} or
{\it available work\/} upon observing state $j$ can now be written as
\begin{equation}
\Delta F_{j,\rm tot} =- T\, [\Delta S+k_B\ln2\,I_U(s_j|s)]
= -k_BT\ln2\, [-H + I_U(s_j|s)] \;.  \label{deltafj}
\end{equation}
This definition of total free energy is closely related to Zurek's
definition of physical entropy\cite{Zurek1989b}.  Average
conditional algorithmic information $\overline{I_U(\cdot|s)}=\sum
p_jI_U(s_j|s)$ obeys the double inequality\cite{Zurek1989b,Caves1990c}
\begin{equation}
H \leq \overline{I_U(\cdot|s)} < H+O(1) \;,
\label{infobounds}
\end{equation}
where $O(1)$ denotes a positive computer-dependent
constant\cite{Chaitin1987a}. It follows immediately that the {\it
average\/} change in total free energy, $\Delta F_{\rm tot}=\sum
p_j\Delta F_{j,\rm tot}$, is zero or negative:
\begin{equation}
0 \geq \Delta F_{\rm tot} > -O(1)k_BT\ln2 \;.
\label{febounds}
\end{equation}
The left side of this double inequality establishes that acquiring
information cannot increase available work on the average.  For
standard choices for the universal computer $U$, e.g., a Turing
machine or Chaitin's LISP-based universal computer\cite{Chaitin1987a},
the computer-dependent $O(1)$ constant on the right is completely
negligible in comparison with thermodynamic
entropies. Equation~(\ref{febounds}) therefore expresses that on the
average, with respect to a standard universal computer, total free
energy remains essentially unchanged upon observation. Despite the
success of this theory, the presence of an arbitrary constant is
disturbing. To understand the issues involved in removing the
arbitrary constant, we must introduce the notions of simple and
complex states.

Although the average information $\overline{I_U(\cdot|s)}$ is greater
than or equal to $H$, there is a class of low-entropy states that can
be prepared without gathering a large amount of information. For
example, in order to compress a gas into a fraction of its original
volume, free energy has to be spent, but the length in bits of written
instructions to prepare the compressed state is negligible on the
scale of thermodynamic entropies. States that can be prepared reliably
in a laboratory experiment usually are {\it simple states\/}, which
means that there is a short verbal description of how to prepare such
a state.

The concept of a simple state is formalized in algorithmic information
theory. A simple state is defined as a state for which $I_U(s_j|s)\ll
H$; i.e., descriptions for simple states can be generated by short
programs. The total free energy increases, in the sense of
Eq.~(\ref{deltafj}), upon observing the system to be in a simple
state. Simplicity is a computer-dependent concept. Standard universal
computers like Turing machines reflect our intuitive notion of
simplicity. It is easy, however, to define a universal computer for
which there are no short programs at all; such a computer would not
recognize simplicity.

Intuitively, simplicity ought to be an intrinsic property of a
state. A computer formalizing the intuitive concept of simplicity
should reflect this. In particular, for such a computer a simple state
should have a short program independent of the probability
distribution $p_1,\ldots,p_N$. This is not true for all universal
computers. In Sec.~\ref{opt} we introduce a universal computer
$U_{\epsilon}$ for which $I_{U_{\epsilon}}(s_j|s)$ is determined
solely by the probabilities $p_1,\ldots,p_N$. For this computer, a
short program for the $j$th state reflects a large probability $p_j$,
not an intrinsic property of the state. We will say that such a
computer does not recognize intrinsically simple states.

Simple states are rare---there are fewer than $2^n$ states $j$ for
which $I_U(s_j|s)<n$\cite{Chaitin1987a}---and thus arise rarely as the
result of an observation, yet they are of great importance.  Simple
states are states for which the algorithmic contribution to total free
energy is negligible. The concept of total free energy does not
conflict with conventional thermodynamics because thermodynamic states
are simple. If the theory does not have the notion of simple states,
the connection with conventional thermodynamics is lost.

The opposite of a simple state, a {\it complex state}, is defined as a
state for which $I_U(s_j|s)$ is of the same order as $H$.  Complex
states arise not just through Maxwell demon-like observations. We have
shown\cite{Caves1993b,Schack1992a,Schack1993e,Schack1994b} that
initially simple states of chaotic Hamiltonian systems in the presence
of a perturbing environment rapidly evolve into extremely complex
states\cite{Caves1993a,Caves1993b} for which the negative algorithmic
contribution to total free energy is vastly bigger than $H$ and thus
totally dominates conventional free energy.  In addition to giving
insight into the second law of thermodynamics, this result leads to a
new approach to quantum
chaos\cite{Caves1993b,Schack1993e,Schack1994b}.

In this paper, we show how the computer-dependent $O(1)$ constant can
be eliminated from the theory summarized above. In Sec.~\ref{opt} we
construct an optimal universal computer for which the $O(1)$ constant
is minimal. It turns out, however, that optimal universal computers do
not recognize intrinsically simple states and thus are unsatisfactory
in formulating the theory.  This difficulty is solved
in Sec.~\ref{twobit} where we show that any universal computer $U$ can
be modified in a simple way such that (a) any state that is simple
with respect to $U$ is also simple with respect to the modified
universal computer $U_3$ and (b) average conditional information
with respect to $U_3$ exceeds average conditional information with
respect to an optimal universal computer by at most $0.5$ bits.
Moreover, conditional algorithmic information with respect to the
modified computer $U_3$ obeys the inequality
$H\leq\overline{I_{U_3}(\cdot|s)}<H+1$. This double bound is the
tightest possible in the sense that there is no tighter bound that is
independent of the probabilities $p_j$.

\section{AN OPTIMAL UNIVERSAL COMPUTER}   \label{opt}

The idea of an optimal universal computer is motivated by Zurek's
discussion\cite{Zurek1989b} of {\it Huffman
coding\/}\cite{Huffman1952} as an alternative way to quantify the
information in an observational record. We consider only binary codes,
for which the code words are binary strings. Before reviewing Huffman
coding, we need to formalize the concept of a list consisting of
descriptions of $N$ states together with their probabilities.

\vspace{3mm}\noindent{\bf Definition 1:} A {\it list of states\/} $s$
is a string of the form $s=((s_1,p_1),\ldots,(s_N,p_N))$ where
$N\geq2$, $0<p_1\leq\ldots\leq p_N$, $\sum p_j=1$, and $s_j$ is a
binary string ($j=1,\ldots,N$). More precisely, the list of states $s$
is the binary string obtained from the list
$((s_1,p_1),\ldots,(s_N,p_N))$ by some definite translation scheme.
One possible translation scheme is to represent parentheses, commas,
and numbers (i.e., the probabilities $p_j$) in ascii code, and to
precede each binary string $s_j$ by a number giving its length $|s_j|$
in bits.  The entropy of a list of states is $H(s)=-\sum
p_j\log_2p_j$. Throughout this paper, $|t|$ denotes the length of the
binary string $t$.

\vspace{3mm}
The Huffman code for a list of states $s=((s_1,p_1),\ldots,(s_N,p_N))$
is a prefix-free or instantaneous code\cite{Welsh1988}---i.e., no code
word is a prefix of any other code word---and can, like all
prefix-free codes, be represented by a binary tree as shown in
Fig.~\ref{treehuff}. The
% FIGURE 1
number of links leading from the root of the tree to a node is called
the {\it level\/} of that node.  If the level-$n$ node $a$ is
connected to the level-$(n+1)$ nodes $b$ and $c$, then $a$ is called
the {\it parent\/} of $b$ and $c$\/; $a$'s {\it children\/} $b$ and
$c$ are called {\it siblings}.  There are exactly $N$ terminal nodes
or {\it leaves}, each leaf corresponding to a state $j$. Each link
connecting two nodes is labeled 0 or 1. The sequence of labels
encountered on the path from the root to a leaf is the code word
assigned to the corresponding state. The code-word length of a state is
thus equal to the level of the corresponding leaf. Each node is
assigned a probability $q_k$ such that the probability of a leaf is
equal to the probability $p_j$ of the corresponding state and the
probability of each non-terminal node is equal to the sum of the
probabilities of its children.

A binary tree represents a Huffman code if and only if it has the {\it
sibling property\/}\cite{Gallager1978}, i.e., if and only if each node
except the root has a sibling, and the nodes can be listed in order of
nonincreasing probability with each node being adjacent to its sibling
in the list.  The tree corresponding to a Huffman code and thus the
Huffman code itself can be built recursively. Create a list of $N$
nodes corresponding to the $N$ states. These $N$ nodes will be the
leaves of the tree that will now be constructed. Repeat the following
procedure until the tree is complete: Take two nodes with smallest
probabilities, and make them siblings by generating a node that is
their common parent; replace in the list the two nodes by their
parent; label the two links branching from the new parent node by 0
and 1.

The procedure outlined above does not define a unique Huffman code for
the list of states $s$, nor does it give generally a unique set of
code-word lengths. In the following, we will assume that we are given
some definite algorithm to assign a Huffman code where the freedom in
the coding procedure is used to assign to the first state (the one
with smallest probability) a code word of maximum length consisting
only of zeros.

\vspace{3mm}\noindent{\bf Definition 2:} Given a list of states
$s=((s_1,p_1),\ldots,(s_N,p_N))$, the binary string $c_j(s)$ with
length $l_j(s)\equiv|c_j(s)|$ denotes the Huffman code word assigned
to the $j$th state using a definite algorithm with the property that
$c_1(s)=0\ldots0$ and $l_j(s)\leq l_1(s)$ for $j=2,\ldots,N$. We
denote the average Huffman code-word length by $\bar{l}(s)=\sum
p_jl_j(s)$. The {\it redundancy\/} $r(s)$ of the Huffman code is
defined by $r(s)=\bar{l}(s)-H(s)$.

\vspace{3mm}\noindent
The redundancy $r(s)$ obeys the bounds $0\leq r(s)<1$, corresponding
to bounds
\begin{equation}
H(s)\leq\bar{l}(s)<H(s)+1    \label{meanbounds}
\end{equation}
for the average code-word length.
Huffman coding is optimal in the sense that there is no prefix-free
binary code with an average code-word length less than $\bar{l}(s)$.
There can be, however, optimal prefix-free codes that are not
Huffman codes.

The length $l_j(s)$ of the Huffman code word $c_j(s)$ cannot be
determined from the probability $p_j$ alone, but depends on the entire
set of probabilities $p_1,\ldots,p_N$.  The tightest general bounds
for $l_j(s)$ are\cite{Katona1976}
\begin{equation}
1 \leq l_j(s) < -\log_g p_j+1 \;,
\label{huffbounds}
\end{equation}
where $g=(\sqrt{5}+1)/2$ is the golden mean. The code-word length for
some states $j$ thus can differ widely from the value
$-\log_2p_j$. For most states $j$, however, the Huffman code-word
length is $l_j(s)\simeq-\log_2p_j$. The following
theorem\cite{Schack1994d} is a precise version of this statement.

\vspace{3mm}\noindent{\bf Theorem 1:}
{\bf (a):} $P_m^-=\sum_{j\in I_m^-}p_j<2^{-m}$ where
$I_m^-=\{i\mid l_i(s)<-\log_2p_i-m\}$, i.e., the probability that a state
with probability $p$ has Huffman code-word length smaller than
$-\log_2p-m$ is less than $2^{-m}$. (This is true for any prefix-free
code.)
{\bf (b):} $P_m^+=\sum_{j\in I_m^+}p_j<2^{-c(m-2)+2}$ where
$I_m^+=\{i\mid l_i(s)>-\log_2p_i+m\}$ and
$c=(1-\log_2g)^{-1}-1\simeq2.27$, i.e., the probability that a state
with probability $p$ has Huffman code-word length greater than
$-\log_2p+m$ is less than $2^{-c(m-2)+2}$.

\vspace{3mm}{\it Proof\/}: See\cite{Schack1994d}. \hfill$\Box$

\vspace{3mm}
Suppose that one characterizes the information content of a state $j$
by its Huffman code-word length $l_j(s)$. Then in
Eq.~(\ref{infobounds}) average algorithmic information
$\overline{I_U(\cdot|s)}$ is replaced by average code-word length
$\bar{l}(s)$, the $O(1)$ constant is replaced by 1, and
Eq.~(\ref{febounds}) assumes the concise form $0\geq\Delta F_{\rm
tot}>-k_BT\ln2$.  This way of eliminating the $O(1)$ constant,
however, has a high price. Since Huffman code-word lengths depend
solely on the probabilities $p_1,\ldots,p_N$---states with high
probability are assigned shorter code words than states with low
probability---Huffman coding does not recognize intrinsically simple
states.  This means that one of the most appealing features of the
theory is lost, namely that the Landauer erasure cost associated with
states that can be prepared in a laboratory is negligible.

In the present article, we show that it is possible to retain this
feature of the theory, yet still eliminate the computer-dependent
constant. We first attempt to do this by constructing an optimal
universal computer, i.e., a universal computer for which the $O(1)$
constant in Eq.~(\ref{infobounds}) is minimal. We find, however,
that optimal universal computers do not recognize intrinsically simple
states, either. A solution to this problem will be given in
Sec.~\ref{twobit} where we discuss a class of nearly optimal universal
computers.

We will need precise definitions of a computer and a universal
computer, which we quote from Chapter~6.2 in\cite{Chaitin1987a}.

\vspace{3mm}\noindent{\bf Definition 3:}
A {\it computer\/} $C$ is a computable partial function that carries a
program string $p$ and a free data string $q$ into an output string
$C(p,q)$ with the property that for each $q$ the domain of $C(.,q)$ is
a prefix-free set; i.e., if $C(p,q)$ is defined and $p$ is a proper
prefix of $p'$, then $C(p',q)$ is not defined. In other words,
programs must be self-delimiting.  $U$ is a {\it universal computer\/}
if and only if for each computer $C$ there is a constant ${\rm
sim}(C)$ with the following property: if $C(p,q)$ is defined, then
there is a $p'$ such that $U(p',q)=C(p,q)$ and $|p'|\leq|p|+{\rm
sim}(C)$.

\vspace{3mm}\noindent
In this definition, all strings are binary strings, and $|p|$ denotes
the length of the string $p$ as before. The self-delimiting or
prefix-free property entails that for each free data string $q$, the
set of all valid program strings can be represented by a binary tree.

For any binary string $t$ we denote by $t^*(U)$ (or just $t^*$ if no
confusion is possible) the shortest string for which
$U(t^*,\Lambda)=t$ where $\Lambda$ is the empty string; i.e., $t^*$ is
the shortest program for the universal computer $U$ to calculate
$t$. If there are several such programs, we pick the one that is first
in lexicographic order. This allows us to define conditional
algorithmic information.

\vspace{3mm}\noindent{\bf Definition 4:}
The {\it conditional algorithmic information\/} $I_U(t_1|t_2)$ to
specify the binary string $t_1$, given the binary string $t_2$, is
\begin{equation}
I_U(t_1|t_2)=\min_{p\mid U(p,t^*_2)=t_1}|p| \;.
\end{equation}
In words, $I_U(t_1|t_2)$ is the length of a shortest program for $U$
that computes $t_1$ in the presence of the free data string $t^*_2$.
In particular, the conditional algorithmic information $I_U(s_j|s)$ to
specify the $j$th state, given a list of states
$s=((s_1,p_1),\ldots,(s_N,p_N))$, is
\begin{equation}
I_U(s_j|s)=\min_{p\mid U(p,s^*)=s_j}|p| \;.
\end{equation}
The average of $I_U(s_j|s)$ is denoted by $\overline{I_U(\cdot|s)} =
\sum p_j I_U(s_j|s)$.

\vspace{3mm}
The next theorem puts a lower bound on the average information.

\vspace{3mm}\noindent{\bf Theorem 2:}
For any universal computer $U$ and any list of states
$s=((s_1,p_1),\ldots,(s_N,p_N))$, the average conditional algorithmic
information obeys the bound
\begin{equation}
\overline{I_U(\cdot|s)} \geq H(s) + r(s) + p_1 \;.
\label{computerbound}
\end{equation}

\vspace{3mm}{\it Proof\/}:
We denote by $s'_j$ a shortest string for which $U(s'_j,s^*)=s_j$.
The $N$ strings $s'_j$ form a prefix-free code.  If the $N$ strings
$s'_j$ are represented by the leaves of a binary tree, then there is
at least one node that has no sibling. Otherwise $U(p,s^*)$ would be
defined only for a finite number $N$ of programs $p$, and $U$ would
not be a universal computer.  Let us denote by $\cal Q$ a sibling-free
node and by $q$ its probability ($q\geq p_1$). Then a shorter
prefix-free code $\{s''_j\}$ can be obtained by moving node $\cal Q$
down one level. More precisely, for states $j$ corresponding to leaves
of the subtree branching from node $\cal Q$, $s''_j$ is obtained from
$s'_j$ by removing the digit corresponding to the link between node
$\cal Q$ and its parent; for all other states $j$, $s''_j=s'_j$. The
code-word lengths of the new code are $|s''_j|=|s'_j|-1$ if state $j$
is a leaf of the subtree branching from node $\cal Q$ and
$|s''_j|=|s'_j|$ otherwise. Since the new code is prefix-free, its
average code-word length is greater than or equal to the Huffman
code-word length $\bar l(s)$. It follows that
\begin{equation}
\overline{I_U(\cdot|s)} = \sum_j p_j|s'_j|
= \sum_j p_j|s''_j| + q \geq \bar l(s)+p_1 = H(s)+r(s)+p_1 \;,
\label{proof2}
\end{equation}
which proves the theorem. \hfill$\Box$

\vspace{3mm}
We can now proceed to define an optimal universal computer.

\vspace{3mm}\noindent{\bf Definition 5:}
$U$ is an {\it optimal universal computer\/} if there is a constant
$\epsilon>0$ such that for all lists of states
$s=((s_1,p_1),\ldots,(s_N,p_N))$ with $p_1\geq\epsilon$
the average conditional algorithmic information has its minimum
value
\begin{equation}
\overline{I_U(\cdot|s)} = H(s) + r(s) + p_1 \;.
\label{optcompinfo}
\end{equation}

\vspace{3mm}\noindent{\bf Theorem 3:}
For any $\epsilon>0$ there is an optimal universal computer
$U_\epsilon$.

\vspace{3mm}{\it Proof\/}:
Let $U$ be an arbitrary universal computer and $\epsilon>0$.  For any
list of states $s=((s_1,p_1),\ldots,(s_N,p_N))$ with
$p_1\geq\epsilon$ we define $c'_1(s)=c_1(s)\circ1=0\ldots01$ and
$c'_j(s)=c_j(s)$ for $j=2,\ldots,N$ where $\circ$ denotes
concatenation of strings. The strings $c'_j(s)$ thus differ from the
Huffman code $c_j(s)$ in that a 1 has been appended to the code
word for the state $j=1$.  According to Eq.~(\ref{huffbounds}),
$l_1(s)+1\leq N_0\equiv\lfloor-\log_g\epsilon+2\rfloor$, where
$g=(\sqrt{5}+1)/2$ and $\lfloor x\rfloor$ denotes the largest integer
less than or equal to $x$.  We denote by $\sigma_0$ a string composed
of $N_0$ zeros; none of the strings $c'_j(s)$ is longer than
$\sigma_0$.

For the definition of  $U_{\epsilon}(p,q)$ we distinguish two cases.
If the binary string $q$ is of the form
\begin{equation}
q=\sigma_0\circ q_s \;\;\mbox{ with }\;\; U(q_s,\Lambda)=s
\label{qform}
\end{equation}
for some list of states $s=((s_1,p_1),\ldots,(s_N,p_N))$ with
$p_1\geq\epsilon$, then $U_\epsilon(p,q)$ is defined for
\begin{equation}
p\in D(q)\equiv\{\sigma_0\circ p'\mid U(p',q) \mbox{ is
defined}\}\cup\{c'_j(s)\mid1\leq j\leq N\} \;,
\end{equation}
with
\begin{equation}
U_\epsilon(\sigma_0\circ p',q)=U(p',q) \mbox{~whenever~} U(p',q)
\mbox{~is defined}
\end{equation}
and
\begin{equation}
U_{\epsilon}(c'_j(s),q)=s_j \mbox{~~for~~} j=1,\ldots,N  \;.
\end{equation}
If the binary string $q$ is not of the form~(\ref{qform}),
then $U_\epsilon(p,q)$ is defined for
\begin{equation}
p\in D(q)\equiv\{\sigma_0\circ p'\mid U(p',q) \mbox{ is defined}\}  \;,
\end{equation}
with
\begin{equation}
U_\epsilon(\sigma_0\circ p',q)=U(p',q) \mbox{~whenever~} U(p',q) \mbox{~is
defined} \;.
\end{equation}
In both cases, the set $D(q)$, which is the domain of
$U_\epsilon(\cdot,q)$, is clearly prefix-free. Moreover, since
$U_\epsilon(\sigma_0\circ p,q)=U(p,q)$ whenever $U(p,q)$ is defined
and $U$ is a universal computer, $U_\epsilon$ is also a universal
computer, with the simulation constant ${\rm sim}(C)$ increased by
$N_0$.

For any string $t$ the minimal program on $U_\epsilon$---i.e., the
shortest program given an empty free data string---is $t^*(U_\epsilon)=
\sigma_0\circ t^*(U)$, where $t^*(U)$ is the minimal program for $t$ on
$U$. In particular, the shortest program for $U_\epsilon$ to compute $s$ is
$s^*(U_\epsilon)=\sigma_0\circ s^*(U)$. Since
$U_\epsilon(c'_j(s),s^*(U_\epsilon))=s_j$ and $|c'_j(s)|\leq N_0$ for
$j=1,\ldots,N$ while $|p|\geq N_0$ for all other programs $p\in
D(s^*(U_\epsilon))$, it follows immediately that
\begin{equation}
I_{U_{\epsilon}}(s_j|s)=|c'_j(s)|=|c_j(s)|+\delta_{1j}=
l_j(s)+\delta_{1j}
\end{equation}
and thus that
\begin{equation}
\overline{I_{U_{\epsilon}}(\cdot|s)}=\sum
p_jI_{U_\epsilon}(s_j|s)=\sum p_j|c'_j(s)|=\sum p_j|c_j(s)|+p_1=\bar
l(s)+p_1=H(s)+r(s)+ p_1 \;.
\end{equation}
\hfill$\Box$

\vspace{3mm}
If $U(q_s,\Lambda)=U_{\epsilon}(\sigma_0\circ q_s,\Lambda)=s$, i.e.,
if $q_s$ is a program for $U$ generating a list of states $s$, the
programs $p$ for which $U_{\epsilon}(p,\sigma_0\circ q_s)$ is defined
can be represented by a binary tree similar to Fig.~\ref{treeopt}.
% FIGURE 2
With respect to the binary tree representing the Huffman code
(Fig.~\ref{treehuff}), the leaf for the $j=1$ state has been moved up
one level to make room for the new node labeled by $U$. This new node
leads to a subtree representing all programs $p'$ for which
$U(p',\sigma_0\circ q_s)$ is defined.

The operation of the optimal universal computer $U_\epsilon$ can be
described in the following way.  When $U_\epsilon$ reads a string that
begins with $N_0$ zeros from its program tape, $U_\epsilon$ disregards
the $N_0$ zeros and interprets the rest of the string as a program for
the universal computer $U$, executing it accordingly.  If $U_\epsilon$
encounters the digit 1 while reading the first $N_0$ digits from its
program tape, $U_\epsilon$ interrupts reading from the program tape,
reads in the free data string, and executes it. If the result of
executing the free data string is a list of states
$s=((s_1,p_1),\ldots,(s_N,p_N))$, $U_\epsilon$ establishes the
modified Huffman code $\{c'_j(s)\}$ for $s$, continues reading
digits from the program tape until the string read matches one of the
code words, say $c'_{j_0}(s)$, and then prints the string $s_{j_0}$.
The output of $U_\epsilon$ is undefined in all other cases.

Since $r(s)+p_1<1$\cite{Gallager1978},
$H(s)\leq\overline{I_U(\cdot|s)}<H(s)+1$ for any optimal universal
computer $U$. For the particular optimal universal computer
$U_\epsilon$ defined in the proof of theorem 3, however, the
information $I_{U_\epsilon}(s_j|s)$ is completely determined by the
Huffman code-word length for the $j$th state and therefore is
completely determined by the probabilities $p_1,\ldots,p_N$. This
optimal universal computer does not recognize intrinsically simple
states. As an aside, note that $U_\epsilon$ cannot give a short
description of the background information for any probability
distribution, because a minimal program for computing the list of
states $s$ on $U_\epsilon$ must begin with $N_0$ zeros.
It turns out that all optimal
universal computers, not just $U_\epsilon$, are unable to recognize
intrinsically simple states. The following theorem formulates this
inability for all optimal universal computers in a slightly weaker
form than holds for $U_\epsilon$. As a consequence, the use of
algorithmic information with respect to an optimal universal computer
to quantify the information in an observational record presents no
advantage over the use of Huffman coding.

\vspace{3mm}\noindent{\bf Theorem 4:}
For any optimal universal computer $U$ and any list of
states $s=((s_1,p_1),\ldots,(s_N,p_N))$ for which
$\overline{I_U(\cdot|s)}=H(s)+r(s)+p_1$, the
following holds: If $p_i>p_j$, then $I_U(s_i|s) \leq
I_U(s_j|s)$. Optimal universal computers therefore do not
recognize intrinsically simple states.

\vspace{3mm}{\it Proof\/}:
To prove the theorem, we show that
$\overline{I_U(\cdot|s)}>H(s)+r(s)+p_1$ for any universal computer $U$
and any list of states $s=((s_1,p_1),\ldots,(s_N,p_N))$ for which
there are indices $i$ and $j$ such that $p_i>p_j$ but
$I_U(s_i|s)>I_U(s_j|s)$. We denote by $s'_j$ a shortest string for
which $U(s'_j,s^*)=s_j$. The strings $s'_j$ form a prefix-free
code. Following an argument similar to the proof of theorem 2, we can
shorten that code on the average by moving a sibling-free node one
level down and in addition by interchanging the code words for states
$i$ and $j$.  The resulting shorter code must obey the Huffman bound,
from which the inequality $\overline{I_U(\cdot|s)}>\bar
l(s)+p_1=H(s)+r(s)+p_1$ follows.
\hfill$\Box$

\section{PRESERVING SIMPLE STATES BY GIVING UP 1/2 BIT} \label{twobit}

Although the discussion in the last section shows that optimal
universal computers present no advantages over Huffman coding, the
main idea behind their construction can be further exploited. If the
subtree representing the programs for the universal computer $U$ is
not attached next to the $j=1$ leaf as in Fig.~\ref{treeopt}, but
instead is attached close to the root as in Fig.~\ref{tree3}, the
resulting universal computer $U_3$ combines the desirable properties
of Huffman coding and the computer $U$. This is the content of the
following theorem.

% FIGURE 3

\vspace{3mm}\noindent{\bf Theorem 5:}
For any universal computer $U$ there is a universal computer $U_3$
such that
\begin{equation}
I_{U_3}(t_1|t_2)\leq I_U(t_1|t_2) + 3   \label{3bit}
\end{equation}
for all binary strings $t_1$ and $t_2$, and that
\begin{equation}
H(s)\leq\overline{I_{U_3}(\cdot|s)}<H(s)+1   \label{avhuffbounds}
\end{equation}
and
\begin{equation}
\overline{I_{U_3}(\cdot|s)}\leq H(s)+r(s)+\frac{1}{2}\label{u2-huff}
\end{equation}
for all lists of states $s=((s_1,p_1),\ldots,(s_N,p_N))$.

\vspace{3mm}{\it Proof\/}:
Let $U$ be an arbitrary universal computer.  For any list of states
$s=((s_1,p_1),\ldots,(s_N,p_N))$ we define the set of strings
$c'_j(s)$ as follows. We start from the binary tree formed by the
Huffman code words $c_j(s)$ where we denote by $q_1$ the probability
of the level-1 node connected to the root by the link labeled 0 (see
Fig.~\ref{treehuff}). According to the value of $q_1$, we distinguish
two cases. In the case $q_1\leq1/2$, $c'_j(s)=01\circ c^+_j(s)$ if
$c_j(s)$ is of the form $c_j(s)=0\circ c^+_j(s)$, and $c'_j(s)=c_j(s)$
if $c_j(s)$ is of the form $c_j(s)=1\circ c^+_j(s)$.  In the case
$q_1>1/2$, $c'_j(s)=01\circ c^+_j(s)$ if $c_j(s)$ is of the form
$c_j(s)=1\circ c^+_j(s)$, and $c'_j(s)=1\circ c^+_j(s)$ if $c_j(s)$ is
of the form $c_j(s)=0\circ c^+_j(s)$.

Figure~\ref{tree3} illustrates the binary tree formed by the code
words $c'_j(s)$ for the case $q_1\leq1/2$.  Of the two main subtrees
emerging from the level-1 nodes in Fig.~\ref{treehuff}, the subtree
having smaller probability is moved up one link and attached to the
node labeled 01, and the subtree having larger probability is attached
to the node labeled 1. In this way, the node labeled 00 is freed
for the subtrees representing the valid programs for $U$.

For the definition of $U_3(p,q)$ we distinguish three cases.
If the binary string $q$ is of the form
\begin{equation}
q=000\circ q_s \mbox{~with~} U(q_s,\Lambda)=s
\label{qform2}
\end{equation}
for some list of states $s=((s_1,p_1),\ldots,(s_N,p_N))$, then
$U_3(p,q)$ is defined for
\begin{eqnarray}
\lefteqn{p\in D(q)\equiv }  \nonumber\\
&&\{000\circ p'\mid U(p',q) \mbox{~is defined}\}
\cup\{001\circ p'\mid U(p',q_s) \mbox{~is defined}\}
\cup\{c'_j(s)\mid1\leq j\leq N\} \;,
\end{eqnarray}
with
\begin{equation}
U_3(000\circ p',q)=U(p',q) \mbox{~whenever~} U(p',q)
\mbox{~is defined} \;,
\end{equation}
\begin{equation}
U_3(001\circ p',q)=U(p',q_s) \mbox{~whenever~} U(p',q_s)
\mbox{~is defined} \;,
\end{equation}
and
\begin{equation}
U_3(c'_j(s),q)=s_j \mbox{~for~} j=1,\ldots,N \;.
\end{equation}
If the binary string $q$ is of the form
\begin{equation}
q=000\circ q' \;,
\label{qform3}
\end{equation}
but there is {\it no\/} list of states $s$ such that $U(q',\Lambda)=s$,
then
$U_3(p,q)$ is defined for
\begin{equation}
p\in D(q)\equiv\{000\circ p'\mid U(p',q) \mbox{~is defined}\}
\cup\{001\circ p'\mid U(p',q') \mbox{~is defined}\} \;,
\end{equation}
with
\begin{equation}
U_3(000\circ p',q)=U(p',q) \mbox{~whenever~} U(p',q)
\mbox{~is defined}
\end{equation}
and
\begin{equation}
U_3(001\circ p',q)=U(p',q') \mbox{~whenever~} U(p',q')
\mbox{~is defined} \;.
\end{equation}
Finally, if $q$
is not of the form~(\ref{qform3}), then $U_3(p,q)$ is defined for
\begin{equation}
p\in D(q)\equiv\{000\circ p'\mid U(p',q) \mbox{~is defined}\}  \;,
\end{equation}
with
\begin{equation}
U_3(000\circ p',q)=U(p',q) \mbox{~whenever~} U(p',q)
\mbox{~is defined} \;.
\end{equation}

In all three cases, the set $D(q)$, which is the domain of
$U_3(\cdot,q)$, is clearly prefix-free. Moreover, since $U_3(000\circ
p,q)=U(p,q)$ whenever $U(p,q)$ is defined and $U$ is a universal
computer, $U_3$ is a also a universal computer, with the simulation
constant ${\rm sim}(C)$ increased by 3.
Equation~(\ref{3bit}) holds because of the following.  The minimal
program for $t_2$ on $U_3$ in the presence of an empty free data
string is $t^*_2(U_3)=000\circ t^*_2(U)$ since $U_3(p,\Lambda)$ is
defined only if $p=000\circ p'$ and $U(p',\Lambda)$ is defined, in which
case $U_3(p,\Lambda)=U(p',\Lambda)$.
If $p$ is a minimal program for $t_1$ on
$U$ in the presence of the minimal program for $t_2$, i.e., if
\begin{equation}
U(p,t^*_2(U))=t_1 \;,\; |p|=I_U(t_1|t_2) \;,
\end{equation}
then
\begin{equation}
U_3(001\circ p,t^*_2(U_3))=U_3(001\circ p,000\circ
t^*_2(U))=U(p,t^*_2(U))=t_1
\end{equation}
and therefore
\begin{equation}
I_{U_3}(t_1|t_2)\leq|001\circ p|=|p|+3  \;.
\end{equation}

The strings $c'_j(s)$ form a prefix-free code with an unused code word
of length 2, for which $\sum p_j|c'_j(s)|<H(s)+1$ according to
theorem~3 in\cite{Gallager1978}. (In\cite{Gallager1978}, the
inequality appears with a $\leq$ sign, but equality can occur only if
the smallest probability $p_1$ is equal to zero, a case we have
excluded.)  The shortest program for $U_3$ to compute $s$ is
$s^*(U_3)=000\circ s^*(U)$, where $s^*(U)$ is the shortest program for
$U$ to compute $s$. Since $U_3(c'_j(s),s^*(U_3))=s_j$ for
$j=1,\ldots,N$, it follows immediately that
$I_{U_3}(s_j|s)\leq|c'_j(s)|$ and thus that
\begin{equation}
\overline{I_{U_3}(\cdot|s)}=\sum p_jI_{U_3}(s_j|s)\leq\sum
p_j|c'_j(s)|<H(s)+1,
\end{equation}
which establishes the upper bound in Eq.~(\ref{avhuffbounds}).  The
lower bound in Eq.~(\ref{avhuffbounds}) holds for all universal
computers.  Equation~(\ref{u2-huff}) follows from
\begin{equation}
\sum p_j|c'_j(s)|=\sum p_j|c_j(s)|+\min(q_1,1-q_1)=\bar l(s)+\min(q_1,1-q_1)
\leq H(s)+r(s)+1/2 \;.
\end{equation}
\hfill$\Box$

\vspace{3mm}
If $U(q_s,\Lambda)=U_3(000\circ q_s,\Lambda)=s$, i.e., if $q_s$ is a
program for $U$ generating a list of states $s$, the programs $p$ for
which $U_3(p,000\circ q_s)$ is defined can be represented by a binary
tree similar to Fig.~\ref{tree3}.  The level-3 node labeled $U$ is the
root of a subtree corresponding to the programs $p'$ for which
$U(p',000\circ q_s)$ is defined, and the level-3 node labeled $U'$ is
the root of a subtree corresponding to the programs $p'$ for which
$U(p',q_s)$ is defined.

The operation of the universal computer $U_3$ can be described in the
following way.  When $U_3$ reads a string that begins with the prefix
000 from its program tape, $U_3$ disregards the prefix and interprets
the rest of the string as a program for the universal computer $U$,
executing it accordingly. When $U_3$ reads a string that begins with
the prefix 001 from its program tape, the output is only defined if
the free data string begins with 000, in which case $U_3$ disregards
the first 3 digits of the program and free data strings
and interprets the
rest of the strings as program and free data strings for the universal
computer $U$, executing it accordingly.  If $U_3$ encounters the digit
1 while reading the first two digits from its program tape, $U_3$
interrupts reading from the program tape, reads in the free data
string, and executes it. If the result of executing the free data
string is a list of states $s=((s_1,p_1),\ldots,(s_N,p_N))$, $U_3$
establishes the modified Huffman code $\{c'_j(s)\}$ for $s$,
continues reading digits from the program tape until the string read
matches one of the code words, say $c'_{j_0}(s)$, and then prints the
string $s_{j_0}$.  The output of $U_3$ is undefined in all other
cases.

The computer $U_3$ compromises between the desirable properties of
algorithmic information and Huffman coding. Since algorithmic
information defined with respect to $U_3$ exceeds algorithmic
information relative to $U$ by at most 3 bits, states that are simple
with respect to $U$ are simple with respect to $U_3$.  Those 3 bits
are the price to pay for a small upper bound on average
information. The average conditional algorithmic information
$\overline{I_{U_3}(\cdot|s)}$ obeys the close double bound
Eq.~(\ref{avhuffbounds}) and exceeds the Huffman bound $\bar{l}(s)$ by
at most $0.5$ bits. This half bit is the price to pay for the
recognition of intrinsically simple states.

\section{CONCLUSION}

We have shown that any universal computer $U$ can be modified in such
a way that (i) the modified universal computer $U_3$ recognizes the
same intrinsically simple states as $U$ and (ii) average algorithmic
information with respect to $U_3$ obeys the same close double bound as
Huffman coding, $H(s)\leq\overline{I_{U_3}(\cdot|s)}<H(s)+1$.  If for
any choice of a universal computer $U$, total free energy is defined
with respect to the corresponding modified universal computer $U_3$,
i.e., if the change of total free energy due to finding the system in
the $j$th state is $\Delta F_{j,\rm tot}= -k_BT\ln2\,[-H(s)+
I_{U_3}(s_j|s)]$, then the bounds for the average change in total free
energy are given by
\begin{equation}
0\geq\Delta F_{\rm tot}>-k_BT\ln2
\end{equation}
instead of by Eq.~(\ref{febounds}).

This result effectively eliminates the undetermined computer-dependent
constant from applications of algorithmic information theory to
statistical physics.  Except for an unavoidable loss due to the coding
bounded by $k_BT\ln2$, on the average available work is independent of
the information the observer has acquired
about the system, any decrease of the
statistical entropy being balanced by an equal increase in algorithmic
information.

\acknowledgements

The author wishes to thank Carlton M.~Caves for suggesting the problem
and for many enlightening discussions.

\begin{figure}
%\epsffile{Figtreehuff.ps}
\caption{Binary tree representing the Huffman code for 6 states with
probabilities $p_1,\ldots,p_6$. The node probabilities $q_k$ are
defined recursively, i.e., $q_7=p_1$, $q_8=p_2$, $q_3=q_7+q_8$,
etc. Code words correspond to branch labels; e.g., the
code word for the third state (probability $p_3$) is 110.}
\label{treehuff}
\end{figure}

\begin{figure}
%\epsffile{Figtreeopt.ps}
\caption{Binary tree representing all valid programs  for the
optimal universal computer $U_\epsilon$ in the presence of a free data
string generating a list of states $((s_1,p_1),\ldots,(s_6,p_6))$.
With respect to  the tree in
Fig.~\protect\ref{treehuff}, the node labeled
$q_7=p_1$ has been moved up one level to make room for the subtree
representing programs for $U$.}
\label{treeopt}
\end{figure}

\begin{figure}
%\epsffile{Figtree3.ps}
\caption{Binary tree representing all valid programs  for the
universal computer $U_3$ in the presence of a free data string
generating a list of states $s=((s_1,p_1),\ldots,(s_6,p_6))$.
With respect to  the tree in Fig.~\protect\ref{treehuff},
the level-1 node labeled $q_1$ has been moved up one level to make
room for the subtrees representing programs for $U$. More precisely,
the binary tree represents the programs $p$ for which $U_3(p,000\circ
q_s)$ is defined if $U_3(000\circ q_s,\Lambda)=s$. The node labeled
$U$ is the root of a subtree corresponding to the programs $p'$ for
which $U(p',000\circ q_s)$ is defined, and the node labeled $U'$ is
the root of a subtree corresponding to the programs $p'$ for which
$U(p',q_s)$ is defined.}
\label{tree3}
\end{figure}

\end{document}